\documentclass{article}
\usepackage{spconf,amsmath,amsfonts,graphicx}
\usepackage{multicol,multirow,color}
\usepackage{mathtools}
\usepackage{colortbl}
\usepackage{booktabs}
\usepackage{tabularx}
\usepackage{balance}
\usepackage{url}
\usepackage{enumitem}
\usepackage{multirow}
\usepackage{xspace}
\usepackage{url}
\usepackage[hidelinks]{hyperref}
\usepackage[labelfont=bf]{caption}
\newcolumntype{K}[1]{>{\centering\arraybackslash}p{#1}}

\newcommand{\NAME}{{M3Dsynth}}

\usepackage[labelfont=bf]{caption}
\newcommand{\ru}{\rule{0mm}{3mm}}

\title{M3DSYNTH: A DATASET OF MEDICAL 3D IMAGES WITH AI-GENERATED LOCAL MANIPULATIONS
}
\name{G. Zingarini, D. Cozzolino, R. Corvi, G. Poggi and L. Verdoliva}
\address{University Federico II of Naples, Italy}

\begin{document}
%
\maketitle
\begin{abstract}
The ability to detect manipulated visual content is becoming increasingly important in many application fields, given the rapid advances in image synthesis methods. 
Of particular concern is the possibility of modifying the content of medical images, altering the resulting diagnoses. Despite its relevance, this issue has received limited attention from the research community. One reason is the lack of large and curated datasets to use for development and benchmarking purposes. Here, we investigate this issue and propose M3Dsynth, a large dataset of manipulated Computed Tomography (CT) lung images.
We create manipulated images by injecting or removing lung cancer nodules in real CT scans, 
using three different methods based on Generative Adversarial Networks (GAN) or Diffusion Models (DM), for a total of 8,577 manipulated samples. 
Experiments show that these images easily fool automated diagnostic tools. 
We also tested several state-of-the-art forensic detectors and demonstrated that, 
once trained on the proposed dataset, they are able to accurately detect and localize manipulated synthetic content,
even when training and test sets are not aligned, showing good generalization ability. 
Dataset and code are publicly available at
\href{https://grip-unina.github.io/M3Dsynth/}{\url{https://grip-unina.github.io/M3Dsynth/}}.
\end{abstract}
\section{Introduction}
\label{sec:intro}

Nowadays, the diagnosis of diseases relies heavily on non-invasive medical imaging techniques, such as Magnetic Resonance Imaging (MRI) and Computed Tomography (CT), 
which can produce high-resolution images of the body's internal organs.
3D medical images are typically stored in secure Picture and Archive Communication System (PACS) servers.
In \cite{mirsky2019ct}, however, it was shown that an attacker could enter the system and use deep learning to modify medical CT scans, injecting or removing lung cancer nodules. 
Such actions may have the purpose of committing insurance fraud, falsifying scientific research data or even have political or terrorism-related purposes. 
Unfortunately, such manipulated images can easily fool automated cancer detectors and even medical experts \cite{mirsky2019ct}. 
This motivates the search for methods that can reliably detect and locate such manipulations

In recent years, there has been intense research on fake image detection \cite{verdoliva2020media}, but most efforts have focused on manipulated faces in video (deepfakes) \cite{roessler2019faceforensics++}. 
Indeed, very limited attention has been paid to detect tampering in medical images, except for some works on misuse of images published in biomedical scientific papers \cite{mandelli2022forensic}, 
despite this being a serious threat in several application scenarios \cite{mangaokar2020jekyll}.
The first studies appear in \cite{mirsky2019ct} where a small dataset of 100 tampered medical images is created by injecting or removing in original images nodules (both malignant and benign) generated by a 3D conditional GAN.
The same dataset has been used for several subsequent studies.
In \cite{solaiyappan2022machine} the performance of various machine learning and deep learning detectors is investigated,
showing that high detection accuracy can be obtained through suitable augmentation and fine-tuning.
In \cite{sharafudeen2022medical}, instead, a 3D Convolutional Neural Network (CNN) model is proposed that can better exploit the tridimensional nature of the CT scans, obtaining also in this case good performance.

\begin{figure}[t!]
\centering
\includegraphics[width=0.95\linewidth,trim=0 60 0 0,page=3]{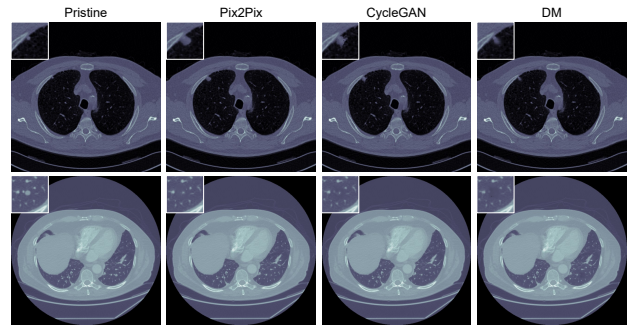}
\caption{Examples of injection (top) and removal (bottom) of a large lung nodule from our dataset. 
Next to the pristine image, we show the manipulated versions obtained from tools based on Pix2Pix, CycleGAN and Diffusion Models.
}
\label{fig:examples}
\end{figure}

\begin{figure*}[t!]
\centering
\includegraphics[width=0.98\linewidth,trim=0 310 0 0,page=2]{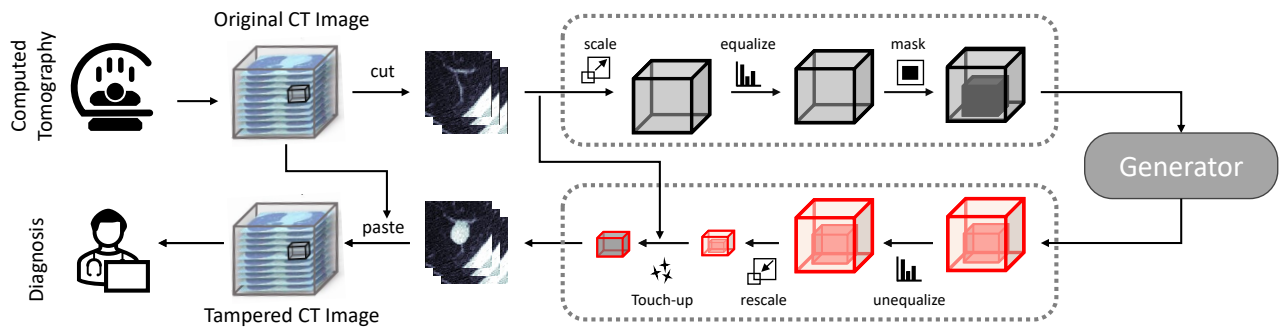}
\caption{Scheme of the manipulation process.
Pre-processing (top): selection of candidate site, extraction of cubic sample, scaling to 32$\times$32$\times$32 pixels, equalization, center masking.
The input datacube feeds a GAN/DM model which generates the synthetic datacube.
Post-processing (bottom): data restoration by de-equalization and inverse scaling, touch-up to improve blending into the host CT scan.
}
\label{fig:pipeline}
\end{figure*}

However, all above detectors only focus on the detection task and were trained and tested on the same type of manipulations, without exploring their ability to generalize.
Indeed, it is well known that data driven detectors perform very well on the image manipulations they were trained for 
but exhibit a sharp performance drop on images displaying new types of synthetic generation \cite{verdoliva2020media}, 
because the forensic traces introduced by unrelated generators may be very different \cite{corvi2023intriguing}.
On the other hand, this latter case is very relevant, since new tools for creating synthetic images are developed by the day.

In this work,
with the aim of providing a solid framework for the development and testing of new forensic detectors,
we propose \NAME, a large dataset of manipulated lung CT images.
Over 8,000 manipulated 3D scans are generated
by injecting or removing lung cancer nodules in real CT scans, using three different GAN/DM generative models.
The size and variety of the proposed dataset allows for a better assessment of new detectors in diverse operating conditions.
We show that the manipulated images easily fool automated diagnosis tools.
Moreover, we carry out a preliminary study of several state-of-the-art detectors
showing that, by training on our large dataset of pristine and synthetic images, 
it is possible to detect and also localize both injections and removals even in a cross-generator scenario.

\section{M3Dsynth Dataset}
\label{sec:dataset}

CT images are precious sources of information for lung cancer diagnosis.
The decision relies mostly on the presence of lung nodules, and especially on their number and size. 
While small nodules are not alarming,
multiple large nodules, with diameter $D>$10mm, represent important diagnostic clues.
Therefore, in our dataset we focus on large (malignant) lung nodules.
Given a real CT scan, we either inject a single malignant nodule in it or remove an existing one from it.
Actually, most of the times we do not create a large nodule anew but enlarge an existing small one and, likewise,
do not remove altogether an existing large nodule but reduce it to a smaller size.
This is to reduce the visual impact of manipulations.
For the same reason, newly generated nodules are placed near existing benign ones.
Fig.\ref{fig:examples} shows a few examples of both manipulations.
The pristine CT scans are from the dataset proposed in \cite{armato2011lung},
comprising 1018 CT scans of 1010 patients, fully annotated with position and size of all detected nodules. Details about training and test sets are present in Tab.\ref{tab:my_label}.

\subsection{Generation}
The manipulation pipeline is described in Fig.\ref{fig:pipeline}.
Only a local 3D cube of the CT scan is modified.
In particular, following \cite{mirsky2019ct}, we consider a cube of side 32mm,
which is much larger than the nodules, both benign and malignant, found in the lungs.
In fact, only the inner part of the cube, of side 16mm, is processed,
masking (zeroing) all input data and generating them anew,
while the outer part is used for conditioning the whole process.
That is, the generator replaces the inner part with synthetic material,
preserving the surrounding anatomical tissue which will therefore fit seamlessly in the original image.
In case of removal, the generator replaces the existing malignant nodule ($D>$10mm) with a smaller nodule ($D<$8mm).
In case of injection, instead, a large nodule ($D>$10mm) is generated.
We manipulate the same CT scan by means of three different generative architectures, two based on GANs and one on DMs.

All networks accept in input a datacube of 32$\times$32$\times$32 pixels.
Note that this requires some preprocessing, 
since a physical cube of side 32mm corresponds to $N$ slices of $M \times M$ pixels in the CT scan, where 
neither $N$ nor $M$ need necessarily be 32, and both numbers depend on the actual CT machine.
Therefore, to obtain a fixed-size network input, the original data are suitably rescaled in advance.
Also, to ensure uniformity, data are also equalized to the same range of values.
Of course, all operations are inverted in output before reinserting the generated data in the original CT scan.

\begin{figure}[t!]
\centering
\includegraphics[width=0.9\linewidth,clip,page=1]{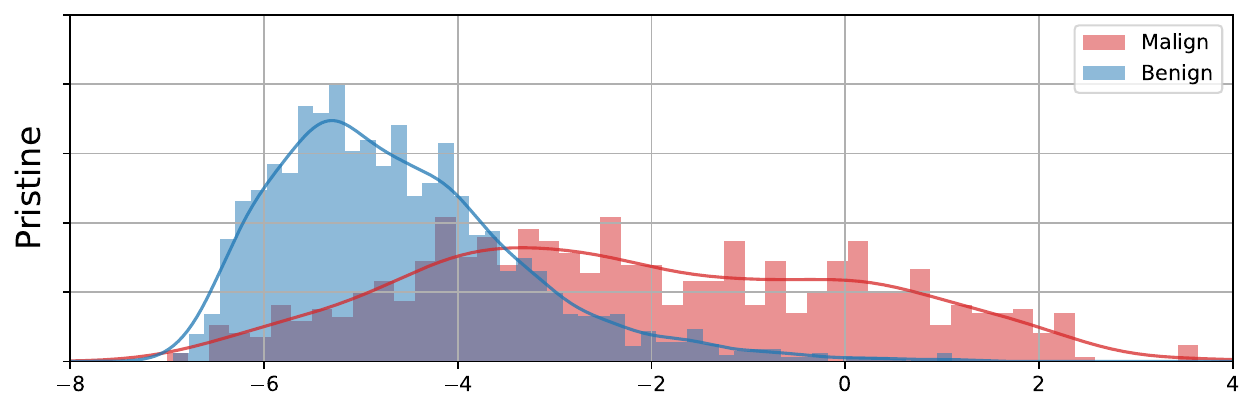}
\includegraphics[width=0.9\linewidth,clip,page=2]{histograms.pdf}

\vspace{-9pt}
\caption{Histograms of lung nodule classification scores.
Top: before manipulation, the diagnostic tools separates relatively well benign (blue) form malignant (red) nodules.
Bottom: after manipulation the removed/shrinked malignant nodules (red) have the same histogram as benign nodules had before manipulation and vice-versa.}
\label{fig:aitool}
\end{figure}

\vspace{2mm}
\noindent
{\bf GAN-based generation.}
Our first generator is the CT-GAN network proposed in \cite{mirsky2019ct},
a 3D version of the Pix2Pix GAN proposed originally \cite{isola2017image} for 2D images.
We train two models of the same architecture, one for the injection and one for the removal task.
Our second generator is based on a 3D CycleGAN, originally designed for the translation of MRI brain images between two different domains \cite{3dcyclemed}.
We adapted it to operate on 3D cubes, considering two translation tasks.
For injection, the translation is between real cancerous tissues and the corresponding masked cubes,
while for removal the translation is between samples without cancer and the corresponding masked cubes.
For the manipulation, we use the two transformations from masked cubes to synthetic cancerous/non-cancerous tissue.

\vspace{2mm}
\noindent
{\bf DM-based generation.}
Finally, we created synthetic samples using a DM tool,
namely, the Denoising Diffusion Probabilistic Model (DDPM) \cite{ho2020denoising},
extended to deal with 3D medical images in \cite{dorjsembe2022three},
by replacing the original denoiser based on a 2D U-Net architecture with an analogous denoiser based on a 3D U-Net.
3D Diffusion Models were also used in \cite{kim2022diffusion} to support the registration of cardiac images.
We adopted the 3D model for our inpainting task, where the generated cube has to be coherent with the available masked cube.
Therefore, with respect to \cite{ho2020denoising,dorjsembe2022three}
we modify the architecture by providing the denoiser with an additional input set to the masked cube.
In detail, we adopted a 2000-step linear scheduling of noise levels and the 3D U-Net-like architecture used in \cite{kim2022diffusion}.

\subsection{Analysis}

To evaluate the quality of generated data, we used the computer-aided diagnostic tool proposed in \cite{liao2019evaluate}.
This is a deep learning-based tool that
comprises a detection network to localize the nodules and a classification network working on each of them.
We applied only the classification network at the position where the nodule was injected or removed.
In Fig.\ref{fig:aitool}, we show the histograms of the classification scores on pristine data (top) and after manipulation (bottom).
Large values indicate malignant nodules.
Although the separation between benign and malignant nodules is not perfect,
in pristine images the classifier tends to assign higher values to malignant nodules\footnote{It is worth pointing out that actual diagnosis for a patient relies on the observation of all nodules in the CT, not just one.}.
After manipulation,
the removed/shrinked nodules have lower scores and the injected/enlarged nodules have larger scores,
with the two histograms exchanging roles.
Therefore, the manipulated data fool the classification tool.

\begin{table}[t]
{\small
\centering
\def\arraystretch{0.6}
\scalebox{0.95}{
\begin{tabular}{lK{1.3cm}K{1.3cm}K{1.3cm}K{1.3cm}} \toprule
           &  Pix2Pix & CycleGAN &     DM &  TOTAL \\ \cmidrule(lr){1-1} \cmidrule(lr){2-4} \cmidrule(lr){5-5}
Injection  &     2009 &     2220 &   2009 &   6238 \\
Removal    &     ~509 &     1016 &   ~814 &   2339 \\ \cmidrule(lr){1-1} \cmidrule(lr){2-4} \cmidrule(lr){5-5}
TOTAL      &     2518 &     3236 &   2823 &   8577 \\
\bottomrule
\end{tabular}
}
\caption{Distribution of manipulated images of the dataset.}
\label{tab:my_label}
}
\end{table}
\begin{table}
{\small
\centering
\scalebox{0.75}{
\setlength{\tabcolsep}{0pt}
\def\arraystretch{1.0}
\begin{tabular}{K{0.4cm}K{1.6cm}|K{1.5cm}K{1.5cm}K{1.5cm}|K{1.5cm}K{1.5cm}K{1.5cm}}
\toprule
\multicolumn{2}{c|}{} & \multicolumn{6}{c}{Test Set} \\
\multicolumn{2}{c|}{ Training} & \multicolumn{3}{c|}{G.P. images}  & \multicolumn{3}{c}{\NAME} \\
\multicolumn{2}{c|}{\ru      Set} & ProGAN & StyleGAN2 & LDM       &  Pix2Pix & CycleGAN &    DM   \\
\cmidrule(lr){1-2} \cmidrule(lr){3-5} \cmidrule(lr){6-8}
\multirow{3}{*}{\rotatebox{90}{\scriptsize G.P. images}}
& \ru ProGAN    & 99.9 & 98.1 & 57.1 &    50.0 &     47.1 &    48.8 \\
& \ru StyleGAN2 & 99.8  & 100 & 57.9 &  50.4 &     49.6 &    52.0 \\
& \ru LDM       & 50.8  & 50.0 & 100 &  44.6 &     44.5 &    46.2 \\
\cmidrule(lr){1-2} \cmidrule(lr){3-5} \cmidrule(lr){6-8}
\multirow{3}{*}{\rotatebox{90}{\scriptsize \NAME}}
& \ru Pix2Pix   & 50.5  & 49.0 & 48.9 &  99.5 &     96.6 &    95.8 \\
& \ru CycleGAN  & 49.5  & 49.0 & 49.9 &  97.7 &     98.5 &    91.6 \\
& \ru DM        & 50.9  & 50.6 & 50.7 &  96.1 &     92.8 &    97.3 \\ \bottomrule
\end{tabular}

}
\caption{Detection Accuracy of the method proposed in \cite{corvi2022detection} (threshold equal to 0.5).
Top: training on general purpose (G.P.) images. Bottom: training on the \NAME\, dataset.
}
\label{tab:comparison}
}
\end{table}

\begin{table*}
{\small
\centering
\def\arraystretch{0.6}
\scalebox{0.95}{
\begin{tabular}{K{0.1cm}lK{1.5cm}K{1.3cm}K{1.3cm}K{1.3cm}K{1.3cm}K{1.3cm}K{1.3cm}K{1.3cm}K{1.3cm}} \toprule
\multicolumn{2}{r}{Test Set:}     &          \multicolumn{3}{c}{Pix2Pix} &         \multicolumn{3}{c}{CycleGAN} &               \multicolumn{3}{c}{DM} \\
\cmidrule(lr){1-2} \cmidrule(lr){3-5} \cmidrule(lr){6-8} \cmidrule(lr){9-11}
\multicolumn{2}{r}{Training Set:} &    Pix2Pix &   CycleGAN &         DM &    Pix2Pix &   CycleGAN &         DM &    Pix2Pix &   CycleGAN &         DM \\ 
\cmidrule(lr){1-2} \cmidrule(lr){3-5} \cmidrule(lr){6-8} \cmidrule(lr){9-11}
\multirow{5}{*}{\rotatebox{90}{F1 ~~/~~IoU }}
 & \ru \cite{ronneberger2015unet} U-Net                         & 44.5 /30.7 & 39.7 /26.6 & 35.5 /23.2 & 34.4 /23.3 & 57.5 /43.6 & 22.7 /15.5 & 46.9 /33.3 & 49.1 /35.8 & 57.7 /43.6 \\
 & \ru \cite{li2019localization} HP-FCN                        & 85.0 /75.3 & 59.1 /43.4 & 45.6 /31.3 & 63.6 /49.8 & 84.5 /75.3 & 36.4 /24.6 & 77.0 /64.9 & 73.6 /61.9 & 84.9 /75.4 \\
 & \ru \cite{wu2019mantra} ManTraNet                     & 87.0 /79.1 & 66.5 /50.5 & 61.4 /45.5 & 74.8 /63.3 & 85.5 /77.2 & 60.5 /47.4 & 83.2 /73.0 & 81.8 /70.7 & 87.2 /78.5 \\
 & \ru \cite{chen2021image} MVSS-Net                      & 81.4 /70.4 & 63.2 /49.8 & 56.8 /42.5 & 74.7 /64.2 & 86.2 /78.0 & 55.1 /44.1 & 79.5 /68.5 & 72.8 /62.2 & 84.9 /75.4 \\
 & \ru \cite{guillaro2023trufor} TruFor                        & 89.9 /82.9 & 68.1 /55.5 & 68.0 /54.7 & 79.0 /70.1 & 88.2 /81.2 & 65.0 /54.1 & 84.4 /75.2 & 76.9 /66.7 & 89.3 /82.0 \\
\cmidrule(lr){1-2} \cmidrule(lr){3-5} \cmidrule(lr){6-8} \cmidrule(lr){9-11}
\multirow{6}{*}{\rotatebox{90}{Acc~~/~~Pd@1\%}}
 & \ru \cite{chollet2017xception} Xception                      & 83.7 /99.8 & 86.9 /95.2 & 71.9 /80.3 & 81.3 /86.1 & 87.4 /99.2 & 64.1 /37.8 & 83.5 /97.7 & 86.8 /94.1 & 71.9 /96.9 \\
 & \ru \cite{ronneberger2015unet} U-Net                         & 52.9 /93.1 & 60.3 /74.5 & 53.7 /56.5 & 52.1 /64.4 & 60.6 /95.4 & 53.0 /29.2 & 52.9 /91.1 & 60.3 /79.5 & 53.7 /96.8 \\
 & \ru \cite{li2019localization} HP-FCN                        & 59.8 /45.6 & 71.4 /50.8 & 60.2 /31.7 & 59.8 /43.1 & 71.4 /52.0 & 60.3 /28.9 & 59.8 /45.4 & 71.4 /51.4 & 60.4 /33.6 \\
 & \ru \cite{wu2019mantra} ManTraNet                     & 52.7 /100. & 56.6 /99.9 & 52.8 /91.2 & 52.7 /93.4 & 56.6 /99.7 & 52.8 /87.3 & 52.7 /99.9 & 56.6 /100. & 52.8 /100. \\
 & \ru \cite{chen2021image} MVSS-Net                      & 73.0 /95.8 & 92.5 /97.2 & 75.4 /86.2 & 72.1 /70.8 & 92.7 /99.3 & 73.7 /67.4 & 73.0 /91.2 & 92.6 /97.9 & 76.0 /99.3 \\
 & \ru \cite{guillaro2023trufor} TruFor                        & 95.0 /100. & 95.8 /97.8 & 94.3 /97.0 & 93.3 /95.9 & 96.0 /99.4 & 91.2 /89.1 & 95.0 /99.9 & 96.0 /98.1 & 94.9 /99.6 \\ \bottomrule
\end{tabular}
}
\caption{Benchmark results on \NAME. 
Top: F1 and IOU localization metrics. Bottom: Accuracy and Pd@1\% detection metrics.}
\label{tab:comparison_det}
}
\vspace{-0.5cm}
\end{table*}

The proposed dataset may be precious to assess new detectors but it is even more important in the training phase.
This latter point can be fully appreciated by means of a further preliminary experiment.
We consider the detector of synthetic images proposed in \cite{corvi2022detection}, 
that is trained general purpose (G.P.) images, and test it on such generic images and on the images of the \NAME~ dataset,
assuming to know in advance the position of the forged area, as the tool performs only detection, not localization.
Results are reported in the upper part of Tab.\ref{tab:comparison}.
On generic images the detector achieves good results, while on medical images 
an accuracy around 50\% is obtained,
showing that the detector, although trained on images generated by both GAN and DM-based methods, has no clue on the nature of test images.
Then, we repeated the analysis after fine tuning on the proposed dataset obtaining very different results (bottom part of Tab.\ref{tab:comparison}).
Now the detection accuracy is always over 90\% even when
training and test images are generated with different methods,
while performance on generic images decreases, as expected.

\section{Benchmark evaluation}
\label{sec:results}
We now test some state-of-the-art forensic methods on the proposed dataset so as to establish a first benchmark.
Not all image forensics tools are appropriate for our task.
In fact, several classical approaches look for compression artifacts or traces of internal camera processing \cite{verdoliva2020media},
but compression is not customary for CT images, and medical imaging sensors have very different properties than smartphones or general-purpose cameras.
Therefore, for the time being,
we restrict attention to deep learning-based methods that may be used or adapted to perform both detection and localization of synthetic content, and such to be easily fine-tuned on our dataset.

\vspace{2mm}
\noindent
{\bf Baseline methods.}
{\bf Xception}
is a generic deep convolutional neural network (CNN) \cite{chollet2017xception}
often used for deepfake detection and as a backbone in more sophisticated architectures \cite{roessler2019faceforensics++}.
Also {\bf U-Net} \cite{ronneberger2015unet}, the well-known segmentation workhorse, is often used as a backbone for multiple forensic tasks \cite{shi2020global, bi2019rru}.
{\bf HP-FCN}
is a fully convolutional network with a high-pass pre-filtering layer, proposed to localize inpainting in natural images \cite{li2019localization}.
{\bf MantraNet}
is a fully convolutional network \cite{wu2019mantra},
that uses a long short-term memory module to assess local anomalies and self-supervised pre-training on 385 different image operations.
{\bf MVSS-Net}
has been proposed recently for image forgery localization and detection in \cite{chen2021image}.
The network has multi-scale modules and two branches to exploit both noise and edge information.
{\bf TruFor} \cite{guillaro2023trufor}
is a transformer-based network that performs forgery localization and detection by
jointly exploiting RGB data and a learned noise-sensitive fingerprint.

All networks are trained using the focal loss for localization, the binary cross-entropy loss for detection, and their combination when both tasks are performed.
When network weights are available, they are imported and fine-tuned on the proposed dataset. 
A pristine CT scan can be used to generate multiple manipulated images by injecting and removing nodules in different places.
Therefore, to avoid any biases, the data are split on a per-patient basis,
with 488 patients for training, 100 for validation and 150 more for testing.
Note that some preliminary experiments have shown that all these methods trained on the original datasets provide unsatisfactory results both for localization and detection on \NAME, hence in the following we fine-tune on our medical data.

\vspace{2mm}
\noindent
{\bf Localization and detection.}
To evaluate their generalization ability, methods are trained on images manipulated by a single synthetic generator and tested against all the others.
Experimental results are collected in Tab.\ref{tab:comparison_det},
where the top part refers to the localization task.
Methods generate a 3D localization map which is compared with the ground truth to compute two performance metrics:
F1 measure and Intersection-over-Union (IoU) score averaged on all images of the same type.
The performance is very good on average, especially for TruFor and ManTraNet, and only plain U-Net seems to struggle.
As expected, the best results are for aligned data (same training and test),
but only a limited impairment is observed on non-aligned data, testifying of a good generalization ability.

In the bottom part, detection results are shown, again for two different metrics.
Accuracy is computed as the (balanced) probability of correct decision.
Decisions are made by comparing the maximum detection score obtained over all slices of an image with the fixed 0.5 threshold.
Several methods provide dismaying results, close to 50\%,
but these may be also due to the fixed choice of the threshold, and could be improved through calibration.
This latter problem is overcome by the second metric Pd@1\%.
In this case, the threshold is set off-line, working only on pristine images, to ensure a 1\% probability of false alarm,
and the corresponding probability of detection measures performance.
A small false alarm rate guarantees correct operations in a real-world scenario where the vast majority of images is pristine.
Despite this strong constraint,
several methods show a very good detection performance and only HP-FCN provides poor results.

\section{Conclusion}
\label{sec:conc}

We introduced \NAME, a large dataset of tampered 3D medical images with local AI-based manipulations that includes CT scans with both injection and removal of lung nodules.
The dataset has been used to train and test several state-of-the-art detectors in different situations and evaluate their performance across different synthetic generators.
Some of the tested methods proved very good both at detecting and localizing local manipulations, even when training and test are mis-aligned.
We hope this dataset will stimulate the research community to work on this topic,
contributing new data and methods,
and exploring challenging situations such as using adversarial attacks to fool the detectors.

\vspace{2mm}
\noindent
{\bf Acknowledgment.}
We gratefully acknowledge the support of this research by the Defense Advanced Research Projects Agency (DARPA) under agreement number FA8750-20-2-1004.
The U.S. Government is authorized to reproduce and distribute reprints for Governmental purposes notwithstanding any copyright notation thereon.
The views and conclusions contained herein are those of the authors and should not be interpreted as necessarily representing the official policies or endorsements, either expressed or implied, of DARPA or the U.S. Government.
This work has also received funding by the European Union under the Horizon Europe vera.ai project, Grant Agreement number 101070093, and is supported by a TUM-IAS Hans Fischer Senior Fellowship and by PREMIER funded by the Italian Ministry of Education, University, and Research within the PRIN 2017 program.

\balance
\bibliographystyle{IEEEbib}
{\small \bibliography{refs}}

\begin{thebibliography}{10}

\bibitem{mirsky2019ct}
Y.~Mirsky, T.~Mahler, I.~Shelef, and Y.~Elovici,
\newblock ``{CT-GAN}: Malicious tampering of 3d medical imagery using deep
  learning,''
\newblock in {\em 28th USENIX Security Symposium}, 2019, pp. 461--478.

\bibitem{verdoliva2020media}
L.~Verdoliva,
\newblock ``{Media Forensics and DeepFakes: an overview},''
\newblock {\em IEEE Journal of Selected Topics in Signal Processing}, vol. 14,
  no. 5, pp. 910 -- 932, 2020.

\bibitem{roessler2019faceforensics++}
A.~R{\"{o}}ssler, D.~Cozzolino, L.~Verdoliva, C.~Riess, J.~Thies, and
  M.~Nie{\ss}ner,
\newblock ``Faceforensics++: Learning to detect manipulated facial images,''
\newblock in {\em IEEE/CVF International Conference on Computer Vision (ICCV)},
  2019, pp. 1--11.

\bibitem{mandelli2022forensic}
S.~Mandelli, D.~Cozzolino, J.~Cardenuto, D.~Moreira, P.~Bestagini, W.~Scheirer,
  A.~Rocha, L.~Verdoliva, S.~Tubaro, and E.~Delp,
\newblock ``Forensic analysis of synthetically generated western blot images,''
\newblock {\em IEEE Access}, vol. 10, pp. 59919 -- 59932, 2022.

\bibitem{mangaokar2020jekyll}
N.~Mangaokar, J.~Pu, P.~Bhattacharya, C.~Reddy, and B.~Viswanath,
\newblock ``Jekyll: Attacking medical image diagnostics using deep generative
  models,''
\newblock in {\em IEEE European Symposium on Security and Privacy (EuroS\&P)},
  2020, pp. 139--157.

\bibitem{solaiyappan2022machine}
S.~Solaiyappan and Y.~Wen,
\newblock ``Machine learning based medical image deepfake detection: A
  comparative study,''
\newblock {\em Machine Learning with Applications}, vol. 8, pp. 1--8, 2022.

\bibitem{sharafudeen2022medical}
M.~Sharafudeen and S.~Chandra,
\newblock ``Medical deepfake detection using 3-dimensional neural learning,''
\newblock in {\em IAPR Workshop on Artificial Neural Networks in Pattern
  Recognition (ANNPR)}, 2022, pp. 169--180.

\bibitem{corvi2023intriguing}
R.~Corvi, D.~Cozzolino, G.~Poggi, K.~Nagano, and L.~Verdoliva,
\newblock ``Intriguing properties of synthetic images: From generative
  adversarial networks to diffusion models,''
\newblock in {\em IEEE/CVF Conference on Computer Vision and Pattern
  Recognition Workshops (CVPRW)}, 2023, pp. 973--982.

\bibitem{armato2011lung}
S.~Armato~III, G.~McLennan, L.~Bidaut, M.~McNitt-Gray, C.~Meyer, A.~Reeves,
  B.~Zhao, D.~Aberle, C.~Henschke, E.~Hoffman, et~al.,
\newblock ``{The lung image database consortium (LIDC) and image database
  resource initiative (IDRI): a completed reference database of lung nodules on
  CT scans},''
\newblock {\em Medical physics}, vol. 38, no. 2, pp. 915--931, 2011.

\bibitem{isola2017image}
P.~Isola, J.-Y. Zhu, T.~Zhou, and A.~Efros,
\newblock ``Image-to-image translation with conditional adversarial networks,''
\newblock in {\em IEEE Conference on Computer Vision and Pattern Recognition
  (CVPR)}, 2017, pp. 1125--1134.

\bibitem{3dcyclemed}
D.~Iommi,
\newblock {\em 3D-CycleGan-Pytorch-Medical-Imaging-Translation},
\newblock \url{https://github.com/davidiommi/3D-CycleGan-Pytorch-MedImaging}.

\bibitem{ho2020denoising}
J.~Ho, A.~Jain, and P.~Abbeel,
\newblock ``Denoising diffusion probabilistic models,''
\newblock {\em Advances in Neural Information Processing Systems (NeurIPS)},
  vol. 33, pp. 6840--6851, 2020.

\bibitem{dorjsembe2022three}
Z.~Dorjsembe, S.~Odonchimed, and F.~Xiao,
\newblock ``Three-dimensional medical image synthesis with denoising diffusion
  probabilistic models,''
\newblock in {\em Medical Imaging with Deep Learning}, 2022, pp. 1--3.

\bibitem{kim2022diffusion}
B.~Kim and J.~C. Ye,
\newblock ``Diffusion deformable model for 4d temporal medical image
  generation,''
\newblock in {\em International Conference on Medical Image Computing and
  Computer-Assisted Intervention (MICCAI)}. Springer, 2022, pp. 539--548.

\bibitem{liao2019evaluate}
F.~Liao, M.~Liang, Z.~Li, X.~Hu, and S.~Song,
\newblock ``{Evaluate the Malignancy of Pulmonary Nodules Using the 3-D Deep
  Leaky Noisy-OR Network},''
\newblock {\em IEEE Transactions on Neural Networks and Learning Systems}, vol.
  30, no. 11, pp. 3484--3495, 2019.

\bibitem{corvi2022detection}
R.~Corvi, D.~Cozzolino, G.~Zingarini, G.~Poggi, K.~Nagano, and L.~Verdoliva,
\newblock ``On the detection of synthetic images generated by diffusion
  models,''
\newblock in {\em IEEE International Conference on Acoustics, Speech and Signal
  Processing (ICASSP)}, 2023, pp. 1--5.

\bibitem{ronneberger2015unet}
O.~Ronneberger, P.~Fischer, and T.~Brox,
\newblock ``U-net: Convolutional networks for biomedical image segmentation,''
\newblock in {\em International Conference on Medical Image Computing and
  Computer-Assisted Intervention (MICCAI)}, 2015, pp. 234--241.

\bibitem{li2019localization}
H.~Li and J.~Huang,
\newblock ``Localization of deep inpainting using high-pass fully convolutional
  network,''
\newblock in {\em IEEE/CVF International Conference on Computer Vision (ICCV)},
  2019, pp. 8301--8310.

\bibitem{wu2019mantra}
Y.~Wu, W.~AbdAlmageed, and P.~Natarajan,
\newblock ``{ManTra-Net: Manipulation Tracing Network for Detection and
  Localization of Image Forgeries With Anomalous Features},''
\newblock in {\em IEEE/CVF Conference on Computer Vision and Pattern
  Recognition (CVPR)}, 2019, pp. 9535--9544.

\bibitem{chen2021image}
X.~Chen, C.~Dong, J.~Ji, J.~Cao, and X.~Li,
\newblock ``{Image Manipulation Detection by Multi-View Multi-Scale
  Supervision},''
\newblock in {\em IEEE/CVF International Conference on Computer Vision (ICCV)},
  2021, pp. 14165--14173.

\bibitem{guillaro2023trufor}
F.~Guillaro, D.~Cozzolino, A.~Sud, N.~Dufour, and L.~Verdoliva,
\newblock ``Trufor: Leveraging all-round clues for trustworthy image forgery
  detection and localization,''
\newblock in {\em IEEE Conference on Computer Vision and Pattern Recognition
  (CVPR)}, 2023, pp. 20606--20615.

\bibitem{chollet2017xception}
F.~Chollet,
\newblock ``Xception: Deep learning with depthwise separable convolutions,''
\newblock in {\em IEEE conference on Computer Vision and Pattern Recognition
  (CVPR)}, 2017, pp. 1251--1258.

\bibitem{shi2020global}
Z.~Shi, X.~Shen, H.~Chen, and Y.~Lyu,
\newblock ``Global semantic consistency network for image manipulation
  detection,''
\newblock {\em IEEE Signal Processing Letters}, vol. 27, pp. 1755--1759, 2020.

\bibitem{bi2019rru}
X.~Bi, Y.~Wei, B.~Xiao, and W.~Li,
\newblock ``{RRU-Net: The Ringed Residual U-Net for Image Splicing Forgery
  Detection},''
\newblock in {\em IEEE/CVF Conference on Computer Vision and Pattern
  Recognition Workshops (CVPRW)}, 2019, pp. 30--39.

\end{thebibliography}

\end{document}